\documentstyle[aaspp4,tighten,flushrt]{article}

\newcommand{\beq}{\begin{equation}}
\newcommand{\eeq}{\end{equation}}
\newcommand{\beqa}{\begin{eqnarray}}
\newcommand{\eeqa}{\end{eqnarray}}

\newcommand{\lexp}{\mathop{\langle}}
\newcommand{\rexp}{\mathop{\rangle}}

\def\d{\delta}
\def\te{\theta}
\def\ds{\delta_s}

\def\dD{[\delta_{\rm D}]}

\def\pl{{ P}}

\font\BF=cmmib10

\def\k{{\hbox{\BF k}}}
\def\x{{\hbox{\BF x}}}
\def\r{{\hbox{\BF r}}}
\def\s{{\hbox{\BF s}}}

\def\v{{\hbox{\BF v}}}
\def\u{{\hbox{$u_z$}}}
\def\uu{{\hbox{\BF u}}}

\def\fun#1#2{\lower3.6pt\vbox{\baselineskip0pt\lineskip.9pt
        \ialign{$\mathsurround=0pt#1\hfill##\hfil$\crcr#2\crcr\sim\crcr}}}

\begin{document}

\hfill{\small CITA-98-16, FERMILAB-Pub-98/254-A}
\normalsize

%
%
\title{The
Bispectrum as a Signature of Gravitational Instability in Redshift-Space}
%
%
\author{Rom\'{a}n Scoccimarro$^{1}$, H.~M.~P.~Couchman$^{2}$,
and Joshua A. Frieman$^{3,4}$}

\vskip 1pc

\affil{${}^1$CITA, McLennan Physical Labs, 60 St George Street,
Toronto, ON M5S 3H8, Canada}

\affil{${}^2$Department of Physics and Astronomy, University of
Western Ontario, London, \\ 
ON N6A 3K7, Canada}

\affil{${}^3$NASA/Fermilab Astrophysics Center, Fermi National
Accelerator Laboratory, \\ P.O. Box 500, Batavia, IL  60510}

\affil{${}^4$Department of Astronomy and Astrophysics,
University of Chicago, Chicago, IL 60637}

\authoremail{scoccima@cita.utoronto.ca}
\authoremail{couchman@uwo.ca}
\authoremail{frieman@fnal.gov}

%
\begin{abstract}

The bispectrum provides a characteristic signature of gravitational
instability that can be used to probe the Gaussianity of the initial
conditions and the bias of the galaxy distribution. We study how this
signature is affected by redshift distortions using perturbation
theory and high-resolution numerical simulations. We obtain
perturbative results for the multipole expansion of the redshift-space
bispectrum which provide a natural way to break the degeneracy between
bias and $\Omega$ present in measurements of the redshift-space power
spectrum.  We propose a phenomenological model that incorporates the
perturbative results and also describes the bispectrum in the
transition to the non-linear regime. We stress the importance of
non-linear effects and show that inaccurate treatment of these can
lead to significant discrepancies in the determination of bias from
galaxy redshift surveys.  At small scales we find that the bispectrum
monopole exhibits a strong configuration dependence that reflects the
velocity dispersion of clusters. Therefore, the hierarchical model for
the three-point function does not hold in redshift-space.

\end{abstract}


\clearpage

%
%
\section{Introduction}
%
%

Understanding the origin of the large-scale distribution of galaxies
in the Universe remains a fundamental goal of cosmology. Galaxy
redshift surveys provide a unique resource for this
endeavor. Recession velocities, however, are not exact indicators of
galaxy radial distances, since gravitational instability induces
peculiar velocities which contribute to galaxy redshifts. This
departure from the Hubble flow causes the observed clustering
distribution in redshift surveys to be statistically anisotropic. On
large scales, the clustering pattern is compressed along the
line-of-sight (the ``squashing effect'') by coherent infall onto
galaxy clusters and superclusters and outflow from voids and other
underdense regions (Sargent \& Turner 1977). On small scales, the
velocity dispersion of virialized clusters creates the so-called
``fingers of god'' (FOG) by stretching out these structures along the
line-of-sight (Jackson 1972). The anisotropy of galaxy redshift-space
clustering provides valuable information about the galaxy velocity
field and hence the underlying mass density field.  In particular,
measurements of redshift-space distortions can be used to determine
the matter density parameter~$\Omega$.

Modelling the redshift distortions of clustering statistics is a
prerequisite for extracting information from redshift surveys (see
Hamilton 1997 for a review). The effects of redshift distortions are
well understood in linear perturbation theory: galaxy power spectrum
measurements on large scales can be used to determine the combination
$\beta \equiv \Omega^{0.6}/b$, where $b$ is the linear bias factor
(Kaiser 1987, Hamilton 1992).  This degeneracy between bias and
$\Omega$ can be lifted in principle by using higher-order statistics,
such as the bispectrum, to independently constrain $b$ (Fry \&
Gazta\~{n}aga 1993; Fry 1994; Gazta\~{n}aga \& Frieman 1994;
Matarrese, Verde \& Heavens 1997). However, to apply this technique to
redshift surveys, redshift distortions of the bispectrum must be taken
into account as well. Hivon et al. (1995) used second-order Lagrangian
perturbation theory to calculate redshift distortions of the
bispectrum monopole, and found them to be rather small, at most
10\%. In this paper, we compute the redshift distortions in Eulerian
perturbation theory and extend their results to the bispectrum
quadrupole. The latter contains valuable information about $\Omega$,
which complements that encoded in the power spectrum distortions. 

As has become evident in recent years from comparisons with numerical
simulations, leading-order perturbative calculations---such as linear
theory for the power spectrum---break down at larger scales in
redshift space than in real space.  In this work, we find a similar
result for the bispectrum. Moreover, mildly non-linear effects are
inevitably present in observations and must be adequately taken into
account to take full advantage of the data (Cole, Fisher \& Weinberg
1994, 1995; Hatton \& Cole 1998). In particular, the next generation
of galaxy redshift surveys, such as the Two Degree Field and Sloan
Digital Sky Surveys, will provide their most precise measurements of
redshift distortions on intermediate scales, where non-linear effects
cannot be neglected. In order to improve upon the perturbative
description of the redshift-space bispectrum, we develop a simple
phenomenological model that extends the perturbative results to
smaller scales, analogous to what has been done for the power spectrum
multipoles.  We compare our results with numerical simulations and
find very good agreement even well into the non-linear regime. We find
that the reduced redshift-space bispectrum $Q_s$ maintains its
configuration dependence at all scales (in contrast to what happens in
real space), due to anisotropies arising from the velocity dispersion
along the line-of-sight. This effect, if not included, would otherwise
compromise the program to use the configuration dependence of the
bispectrum to constrain the bias from measurements in redshift
surveys.

The rest of this paper is organized as follows. In Section~2, we
derive a fully non-linear expression for the density field in redshift
space, which is then used to obtain the perturbative expansion order
by order.  Section~3 applies these results to yield the monopole and
quadrupole moments of the redshift-space bispectrum for different
configurations, including non-linear biasing. The extension of these
results into the non-linear regime via a phenomenological model is the
subject of Section~4, where comparison with numerical simulations is
made. Section 5 provides a final discussion and conclusions.

%
%
\section{Redshift-Space Mapping and Perturbation Theory}
%
%
In redshift space, the radial coordinate $\s$ of a galaxy is given by
its observed radial velocity, a combination of its Hubble flow plus
``distortions'' due to peculiar velocities. The mapping from
real-space position ${\bf \x}$ to redshift space is given by: 
\beq
\s=\x - f \ \u(\x) {\hat z}, 
\eeq 
where $f(\Omega) \approx
\Omega^{0.6}$ is the logarithmic growth rate of linear perturbations,
and $\uu(\x) \equiv - \v(\x)/({\cal H} f)$, where $\v(\x)$ is the
peculiar velocity field, and ${\cal H}(\tau) \equiv (1/a)(da/d\tau)=
Ha$ is the conformal Hubble parameter (with FRW scale factor $a(\tau)$
and conformal time $\tau$).  In Eq.~(1), we have assumed the
``plane-parallel'' (or infinitely remote observer) approximation, so
that the line-of-sight is taken as a fixed direction, denoted by
${\hat z}$.

The density field in redshift space, $\ds(\s)$, is obtained from the
real-space density field $\d(\x)$ by requiring that the redshift-space
mapping conserves the number of galaxies, i.e.

\beq
(1+\ds)d^3s=(1+\d)d^3x \label{mass}~.
\eeq
Using the fact that $d^3s=J(\x)d^3x$, where $J(\x)=1-f \nabla_z
\u(\x)$ is the {\em exact} Jacobian of the mapping in the
plane-parallel approximation, we obtain:

\beq
\ds(\s)= \frac{\d(\x)+1-J(\x)}{J(\x)}= \frac{\d(\x)+f \nabla_z
\u(\x)}{1-f \nabla_z \u(\x)}~.
\eeq
The zeros of the Jacobian describe caustics in redshift space, the
locus of points where the density field is apparently infinite (McGill
1990). This surface is characterized in real space by those points
which are undergoing turn-around in the gravitational collapse
process, so their peculiar velocities exactly cancel the differential
Hubble flow. In practice, caustics are smoothed out by subclustering.

We are interested in the Fourier transform of the density field
contrast in redshift space,

\beq
\ds(\k) \equiv \int \frac{d^3s}{(2\pi)^3} {\rm e}^{-i \k\cdot\s} \ds(\s)
= \int  \frac{d^3x}{(2\pi)^3} {\rm e}^{-i \k\cdot\x}
{\rm e}^{i f k_z \u(\x)} \Big[ \d(\x) + f \nabla_z \u(\x) \Big]\label{d_s}~.
\eeq
This equation describes the fully non-linear density field in redshift
space in the plane-parallel approximation and is the starting point
for the perturbative approach.  The term in square brackets describes
the ``squashing effect'', {\it i.e.}, the increase in the clustering
amplitude due to infall, whereas the exponential factor encodes the
FOG effect, which erases power due to velocity dispersion along the
line-of-sight.  To obtain the perturbative expansion, we expand the
second exponential in power series,

\beq
\ds(\k)=\sum_{n=1}^{\infty} \int d^3k_1 \ldots d^3 k_n
\dD_n \Big[
\d(\k_1) + f \mu_1^2 \te(\k_1) \Big] \frac{ (f \mu
k)^{n-1}}{(n-1)!} \frac{\mu_2}{k_2}\te(\k_2) \ldots
\frac{\mu_n}{k_n}\te(\k_n) \label{delta_s},
\eeq
where $\dD_n \equiv \delta_{\rm D}(\k - \k_1 - \cdots - \k_n)$, with
$\delta_{\rm D}(\x)$ the Dirac delta distribution, the velocity
divergence $\te(\x) \equiv \nabla \cdot \uu (\x)$, and $\mu_i \equiv
\k_i \cdot {\hat z}/k_i$ is the cosine of the angle between the
line-of-sight and the wavevector. In linear perturbation theory, only
the $n=1$ term survives, and we recover the well known formula due to
Kaiser (1987)
\beq
\ds(\k)=\d(\k) (1+f\mu^2) \label{delta_sl}.
\eeq

Equation (\ref{delta_s}) can be used to obtain the redshift-space
density field beyond linear theory. For this purpose, we recall that
the perturbative expansion for the {\it un-redshifted} density and
velocity-divergence fields can be written (Fry 1984, Goroff et
al. 1986)

\beq
\delta(\k,\tau) = \sum_{n=1}^{\infty} D_1^n(\tau)
\int d^3k_1 \ldots \int d^3k_n
\dD_n \, F_n (\k_1, \ldots, \k_n)
	\, \delta_1(\k_1) \cdots \delta_1(\k_n)  ,
\eeq
\beq 
\theta(\k,\tau) = \sum_{n=1}^{\infty} D_1^n(\tau) \int d^3k_1
\ldots \int d^3k_n \dD_n \, G_n (\k_1, \ldots, \k_n) \, \delta_1(\k_1)
\cdots \delta_1(\k_n) , 
\eeq 
where $D_1(\tau)$ is the density perturbation growth factor in linear
theory. Here we have assumed that the n$^{\rm th}$-order growth factor
$D_n \propto D_1^n$, which is an excellent approximation (see
Scoccimarro et al. 1998, Appendix B.3).  The perturbative kernels
$F_n$ and $G_n$ are dimensionless, symmetric, scalar functions of the
wave vectors $ \{ \k_1, \ldots , \k_n \} $ which can be recursively
found from the equations of motion.  Similarly, in redshift space we
can write the density field amplitude in the form

\beq
\delta_s(\k,\tau) = \sum_{n=1}^{\infty} D_1^n(\tau)
\int d^3k_1 \ldots \int d^3k_n
\dD_n \, Z_n (\k_1, \ldots, \k_n)
	\, \delta_1(\k_1) \cdots \delta_1(\k_n)  \label{ec:deltan}.
\eeq

To proceed, we assume a local, non-linear biasing scheme, in which the
galaxy density contrast $\d_g$ is given as a Taylor series expansion
in the underlying dark matter fluctuations $\d$ (Fry \& Gazta\~naga
1993),

\beq
\label{biasing}
\d_g= \sum_{m=0} \frac{b_m}{m!}\  \d^m \equiv b\ (\d + \gamma \d^2/2 +
\ldots )~,
\eeq
where $\gamma=b_2/b$, and $b\equiv b_{1}$ is the linear bias.  If the
galaxy distribution is unbiased, then $b=1$ and $b_m=0$ for $m>1$. The
term $b_0$ ensures that $\langle \d_g \rangle =0$; since it does not
enter into connected correlations, we will neglect it in what follows,
as assumed in the last expression in Eq.~(\ref{biasing}). We note that
the model in Eq.~(\ref{biasing}) is a rather simple bias prescription,
and one could generalize the results below to more complex biasing
schemes (e.g., non-local, time-dependent, stochastic, etc).

From Eqs.~ (\ref{delta_s})-(\ref{biasing}), we can now read off
the redshift-space kernels $Z_n$ for the {\it galaxy} density field.
In particular the linear, second- and third-order perturbation
theory (PT) kernels are

\beqa 
Z_1 (\k) &=& (b+f \mu^2), \label{z1} \\ 
Z_2 (\k_1,\k_2) &=& b F_2 (\k_1,\k_2) +f \mu^2 G_2 (\k_1,\k_2) +
\frac{f \mu k}{2} \Big[ \frac{\mu_1}{k_1} (b+f \mu_2 ^2) +
\frac{\mu_2}{k_2} (b+f \mu_1 ^2) \Big] + \frac{b_2}{2}, \label{z2} \\
Z_3(\k_1,\k_2,\k_3) &=& b F_3^{(s)} (\k_1,\k_2,\k_3) +f \mu^2
G_3^{(s)} (\k_1,\k_2,\k_3) + f \mu k \Big[b F_2^{(s)} (\k_1,\k_2) +f
\mu_{12}^2 G_2^{(s)} (\k_1,\k_2) \Big] \frac{\mu_3}{k_3} \nonumber \\
& & + f \mu k (b+f \mu_1 ^2) \frac{\mu_{23}}{k_{23}}
G_2^{(s)}(\k_2,\k_3) + \frac{(f \mu k)^2}{2} (b+f \mu_1 ^2)
\frac{\mu_2}{k_2} \frac{\mu_3}{k_3} + 3 b_2 F_2^{(s)} (\k_1,\k_2) +
\frac{b_3}{6}, \nonumber \\
\label{z3}
\eeqa 
where we denote $\mu \equiv \k \cdot {\hat z}/k$, with $\k \equiv \k_1
+\ldots +\k_n$, and $\mu_i \equiv \k_i \cdot {\hat z}/k_i$. As above,
$F_2$ and $G_2$ (given in Eqs.~(\ref{f2}-\ref{g2}) below) denote the
second-order kernels for the real-space density and
velocity-divergence fields, and similarly for $F_3$ and $G_3$. Note
that the third order kernel $Z_3$ must still be symmetrized over its
arguments. One can similarly obtain the PT kernels $Z_n$ in redshift
space to arbitrary higher order.  

We note that there are {\em two} approximations involved in this
procedure: one is the mathematical step of going from Eq.~(\ref{d_s})
to Eq.~(\ref{delta_s}), which approximates the redshift-space mapping
with a power series; the other is the PT expansion itself (i.e., the
expansion of $\d(\k)$ and $\te(\k)$ in terms of linear fluctuations
$\d_1(\k)$). Therefore, one is not guaranteed that the resulting PT in
redshift space will work over the same range of scales as in real
space. In fact, we will find that, in general, {\em PT in redshift
space breaks down at larger scales than in real space}, because the
redshift-space mapping is only treated approximately, and it breaks
down at larger scales than does the perturbative dynamics. In
particular, a calculation of the one-loop power spectrum in redshift
space using Eqs.~(\ref{z1}-\ref{z3}) does not give satisfactory
results. To extend the leading-order calculations such as those in
this work, one must treat the redshift-space mapping exactly and only
approximate the dynamics using PT (Scoccimarro, Couchman \& Frieman
1998; hereafter SCF).

The calculation of redshift-space statistics in Fourier space proceeds
along the same lines as in the un-redshifted case. To leading (linear)
order, the redshift-space galaxy power spectrum reads (Kaiser 1987)

\beq
P_s(\k)= P_g(k)\ (1+\beta \mu^2)^2 = \sum_{\ell=0}^{\infty} a_\ell\
\pl_\ell(\mu) \ P_g(k)
\label{mult_l},
\eeq
where $P_g(k)\equiv b^2 P(k)$ is the real-space galaxy power spectrum, $P(k)$
is
the linear mass power spectrum, and $\beta\equiv f/b \approx
\Omega^{0.6}/b$. Here $\pl_\ell(\mu)$ denotes the Legendre polynomial
of order $\ell$, and the multipole coefficients are (Hamilton 1992;
Cole, Fisher \& Weinberg 1994)

\beq
a_0 \equiv 1+ \frac{2}{3}\beta+\frac{1}{5}\beta^2, \ \ \ \ \
a_2 \equiv \frac{4}{3}\beta+\frac{4}{7}\beta^2, \ \ \ \ \
a_4 \equiv \frac{8}{35}\beta^2
\label{PS_l};
\eeq
all other multipoles vanish. Equation~(\ref{mult_l}) is the standard
tool for measuring $\Omega$ from redshift distortions of the power
spectrum in the linear regime; in particular, the
quadrupole-to-monopole ratio $R_{\rm P}\equiv a_2/a_0$ should be a
constant, independent of wavevector $k$, as $k \to 0$. Note, however,
that in these expressions $\Omega$ appears only through the parameter
$\beta$, so there is a degeneracy between $\Omega$ and the linear bias
factor $b$ (it is impossible to distinguish an unbiased low-density
Universe from a biased high-density model).

Given the second-order PT kernel in redshift-space, the leading-order
(tree-level) galaxy bispectrum in redshift-space reads

\beq
\label{Bs_def}
B_s(\k_1,\k_2,\k_3)= 2 Z_2 (\k_1,\k_2) \ Z_1(\k_1)\
Z_1(\k_2) \ P(k_1) \ P(k_2) +
{\rm cyc.},
\eeq
which can be normalized by the power spectrum monopole to give the
hierarchical three-point amplitude in redshift space, $Q_s$,

\beq
Q_s(\k_1,\k_2,\k_3) \equiv \frac{B_s(\k_1,\k_2,\k_3)}{P_s(k_1) \
P_s(k_2) + {\rm cyc.}}
= \frac{B_s(\k_1,\k_2,\k_3)}{a_0^2 \ (P_g(k_1) \
P_g(k_2) + {\rm cyc.})},
\eeq
where ``cyc.'' denotes a sum over permutations of
$\{k_1,k_2,k_3\}$. Note that the hierarchical amplitude $Q_s$ is
independent of power spectrum normalization to leading order in PT.
Since, to leading order, $Q_s$ is a {\em function} of triangle
configuration which separately depends on $\Omega$, $b$, and $b_2$, it
allows one in principle to break the degeneracy between $\Omega$ and
$b$ present in measurement of the power spectrum multipoles in
redshift space.

%
%
\section{Results: The Redshift-Space Bispectrum in Perturbation Theory}
%
%

Due to redshift distortions, the bispectrum becomes a function of five
variables: three of them describe the shape of the triangle (e.g., the
sides $k_1$, $k_2$, and the angle $\theta$ between them, with $\cos \theta
\equiv \hat{\k}_1 \cdot \hat{\k}_2$), and the two
remaining variables characterize the orientation of the triangle with
respect to the line-of-sight, which we take to be the polar angle
$\omega = \cos^{-1} \mu$ of
$\k_1$, and the azimuthal angle $\phi$ about $\hat{\k}_1$:

\beq
\mu_1=\mu = \cos\omega = \hat{\k}_1 \cdot \hat{z}, \ \ \ \ \ \mu_2=\mu \cos
\theta - \sqrt{(1-\mu^2)}\ \sin
\theta\ \cos \phi, \ \ \ \ \ \mu_3= -\frac{k_1}{k_3} \mu -
\frac{k_2}{k_3} \mu_2.
\eeq
The ($\mu$,$\phi$)
dependence introduced by redshift distortions can be conveniently
described by decomposing the tree-level bispectrum $B_s$ in
Eq.~(\ref{Bs_def}) into spherical harmonics,

\beq
B_s(\k_1,\k_2,\k_3)= \sum_{\ell=0}^\infty\ \sum_{m=-\ell}^{\ell}\
B_s^{(\ell,m)}(k_1,k_2,\theta)\ {\rm Y}_{\ell m}(\omega,\phi).
\eeq
Bispectrum multipoles $B_s^{(\ell,m)}$ are non-vanishing
for even $\ell$, up to $\ell=8$, $m=6$. We now explore the
information encoded in these multipoles for different triangle configurations.

%
\subsection{Non-Equilateral Configurations}
%

Rather than working with the full multipole decomposition $(\ell,m)$,
here we concentrate for simplicity on the $m=0$ multipoles. In this
case, which corresponds to averaging over $\phi$, it is more convenient to
decompose the bispectrum in Legendre polynomials

\beq
B_s(\k_1,\k_2,\k_3)= \sum_{\ell=0}^\infty\
B_s^{(\ell)}(k_1,k_2,\theta)\ P_{\ell}(\mu).
\eeq
We can write the different contributions as
\beq
\label{Bst}
B_s^{(\ell)}= P_g(k_1) P_g(k_2)\ \Big( F_2(\k_1,\k_2)\  {\cal D}_{\rm
SQ1}^{\ell} + G_2(\k_1,\k_2)\ {\cal D}_{\rm SQ2}^{\ell} + {\cal D}_{\rm
NLB}^{\ell} + {\cal D}_{\rm FOG}^{\ell}\Big) + {\rm cyc.},
\eeq

\noindent where the second-order real-space kernels are given by
\beq
F_2(\k_1,\k_2)= \frac{5}{7} + \frac{x}{2}\left({k_1 \over k_2}
+ {k_2 \over k_1}\right) + \frac{2x^2 }{7}  \label{f2},
\eeq
\beq
G_2(\k_1,\k_2)= \frac{3}{7} + \frac{x}{2}\left({k_1 \over k_2} +
{k_2 \over k_1}\right) + \frac{4x^2}{7}
\label{g2},
\eeq
with $x\equiv \cos\theta$. The first term in Eq.~(\ref{Bst})
represents the effect of the linear squashing (SQ1), the next term
describes second-order squashing (SQ2), the third contribution is due
to non-linear biasing (NLB), and the last term describes the effect of
damping due to velocity dispersion, the FOG effect.  The different
contributions to the monopole ($\ell=0$) and quadrupole ($\ell=2$) are
given by (recall that $\beta \equiv f/b$, $\gamma=b_2/b$)

\beq
{\cal D}_{\rm SQ1}^{\ell=0} \equiv
{{{2}\,\left( 15 + 10\,\beta  + {{\beta }^2} +
       2\,{{\beta }^2}\,{x^2} \right) }\over {15\, b}},
\eeq
\beq
{\cal D}_{\rm SQ1}^{\ell=2} \equiv
{{2\,\beta \,\left( 7 + \beta  + 21\,{x^2} +
       11\,\beta \,{x^2} \right) }\over {21\, b}},
\eeq
\beqa
{\cal D}_{\rm SQ2}^{\ell=0} &\equiv&
2\,\beta \, ( 35\,{{k_1}^2} +
       28\,\beta \,{{k_1}^2} + 3\,{{\beta }^2}\,{{k_1}^2} +
       35\,{{k_2}^2} + 28\,\beta \,{{k_2}^2} +
       3\,{{\beta }^2}\,{{k_2}^2} + 70\,k_1\,k_2\,x +
       84\,\beta \,k_1\,k_2\,x +\nonumber \\
  & &
       18\,{{\beta }^2}\,k_1\,k_2\,x +
       14\,\beta \,{{k_1}^2}\,{x^2} +
       12\,{{\beta }^2}\,{{k_1}^2}\,{x^2} +
       14\,\beta \,{{k_2}^2}\,{x^2} +
       12\,{{\beta }^2}\,{{k_2}^2}\,{x^2} +
       12\,{{\beta }^2}\,k_1\,k_2\,{x^3} )\nonumber \\
  & &
   /(105\,{{k_3}^2\,  b})
\eeqa

\beqa
{\cal D}_{\rm SQ2}^{\ell=2} &\equiv&
2\,\beta \, ( 14\,{{k_1}^2} +
       13\,\beta \,{{k_1}^2} + {{\beta }^2}\,{{k_1}^2} -
       7\,{{k_2}^2} - 5\,\beta \,{{k_2}^2} +
       28\,k_1\,k_2\,x + 30\,\beta \,k_1\,k_2\,x +
       6\,{{\beta }^2}\,k_1\,k_2\,x + \nonumber \\
  & &
       11\,\beta \,{{k_1}^2}\,{x^2} +
       9\,{{\beta }^2}\,{{k_1}^2}\,{x^2} +
       21\,{{k_2}^2}\,{x^2} + 29\,\beta \,{{k_2}^2}\,{x^2} +
       6\,{{\beta }^2}\,{{k_2}^2}\,{x^2} +
       18\,\beta \,k_1\,k_2\,{x^3} +
       14\,{{\beta }^2}\,k_1\,k_2\,{x^3} + \nonumber \\
  & &
       4\,{{\beta }^2}\,{{k_2}^2}\,{x^4} )
   /(21\,{{k_3}^2}\, b)
\eeqa

\beq
{\cal D}_{\rm NLB}^{\ell=0} \equiv
{{\gamma \,\left( 15 + 10\,\beta  + {{\beta }^2} +
       2\,{{\beta }^2}\,{x^2} \right) }\over {15}\, b}
\eeq
\beq
{\cal D}_{\rm NLB}^{\ell=2} \equiv
{{\beta \,\gamma \,\left( 7 + \beta  + 21\,{x^2} +
       11\,\beta \,{x^2} \right) }\over {21}\, b}
\eeq

\beqa
{\cal D}_{\rm FOG}^{\ell=0} & \equiv &
\beta \, (210\,k_1\,k_2 + 
       210\,\beta \,k_1\,k_2 + 
       54\,\beta^2\,k_1\,k_2 + 
       6\,\beta^3\,k_1\,k_2 + 105\,k_1^2\,x + 
       189\,\beta \,k_1^2\,x + 
       99\,\beta ^2\,k_1^2\,x + \nonumber \\
  & &
       15\,\beta ^3\,k_1^2\,x + 105\,k_2^2\,x + 
       189\,\beta \,k_2^2\,x + 
       99\,\beta ^2\,k_2^2\,x + 
       15\,\beta ^3\,k_2^2\,x + 
       168\,\beta \,k_1\,k_2\,x^2 + \nonumber \\
  & &
       216\,\beta ^2\,k_1\,k_2\,x^2 + 
       48\,\beta ^3\,k_1\,k_2\,x^2 + 
       36\,\beta ^2\,k_1^2\,x^3 + 
       20\,\beta ^3\,k_1^2\,x^3 + 
       36\,\beta ^2\,k_2^2\,x^3 + 
       20\,\beta ^3\,k_2^2\,x^3 + \nonumber \\
  & &
       16\,\beta ^3\,k_1\,k_2\,x^4 ) /  
   (315\,k_1\,k_2)
\eeqa

\beqa
{\cal D}_{\rm FOG}^{\ell=2} &\equiv&
\,\beta \,( 231\,k_1\,k_2 + 
       330\,\beta \,k_1\,k_2 + 
       99\,{{\beta }^2}\,k_1\,k_2 + 
       12\,{{\beta }^3}\,k_1\,k_2 + 462\,{{k_1}^2}\,x + 
       891\,\beta \,{{k_1}^2}\,x + 
       528\,{{\beta }^2}\,{{k_1}^2}\,x + \nonumber \\
  & & 
       75\,{{\beta }^3}\,{{k_1}^2}\,x + 
	462\,{{k_2}^2}\,x + 
       594\,\beta \,{{k_2}^2}\,x + +
       198\,{{\beta }^2}\,{{k_2}^2}\,x + 
       30\,{{\beta }^3}\,{{k_2}^2}\,x + 
       693\,k_1\,k_2\,{x^2} + + \nonumber \\
  & &
       2046\,\beta \,k_1\,k_2\,{x^2} + 
       1485\,{{\beta }^2}\,k_1\,k_2\,{x^2} 
        276\,{{\beta }^3}\,k_1\,k_2\,{x^2} + 
       297\,\beta \,{{k_1}^2}\,{x^3} +
       462\,{{\beta }^2}\,{{k_1}^2}\,{x^3} + 
       205\,{{\beta }^3}\,{{k_1}^2}\,{x^3} +\nonumber \\
  & & 
       594\,\beta \,{{k_2}^2}\,{x^3} + 
       792\,{{\beta }^2}\,{{k_2}^2}\,{x^3} + 
       190\,{{\beta }^3}\,{{k_2}^2}\,{x^3} + 
       396\,{{\beta }^2}\,k_1\,k_2\,{x^4} + 
       272\,{{\beta }^3}\,k_1\,k_2\,{x^4} + \nonumber \\
  & &
       60\,{{\beta }^3}\,{{k_2}^2}\,{x^5} ) / 
   (693\,k_1\,k_2)
 \eeqa

The result for the monopole without bias, when $b=1$ and $\gamma=0$,
agrees with that given in Hivon et al (1995), where they also include
the (extremely weak) $\Omega$-dependence due to second-order PT growth
factors (i.e., the small deviation from $D_{2} \propto D_{1}^{2}$).
This agreement may seem surprising, since they claim to include the
radial character of redshift distortions, whereas our calculations are
done in the plane-parallel approximation. In fact, although
intermediate expressions in Hivon et al (1995) do preserve the radial
character of redshift distortions in the large-volume approximation
[equivalent to replacing $z$ by $r$ in Eq.~(\ref{d_s})], the angle
averaging is done instead assuming that $\r$ is a fixed direction, that
is, in the plane-parallel approximation.  Our results generalize
theirs to include biasing; more importantly, the result for the
bispectrum quadrupole is new.

Figure~1 illustrates these results. The top left panel shows the
amplification of the bispectrum due to redshift-space distortions,
showing the ratio $A_{B}^{(\ell)}\equiv B_{s}^{(\ell)}/B$, for
$\ell=0$ and $\ell=2$, as a function of angle $\theta$, for
configurations with $k_{1}/k_{2}=2$ and scale-free initial conditions,
$P(k) \propto k^n$ ($n=-2$ solid curves, and $n=0$ dotted curves). The
monopole shows a small ($\leq 20\%$) change of the bispectrum due to
redshift distortions, as first shown by Hivon et al.  (1995). The
quadrupole, however, shows a higher amplification for co-linear
configurations (near $\theta=0, \pi$, where $\k_1 || \k_2$) and a
suppression at equilateral-like configurations. This is to be
expected, since the quadrupole is generated solely by redshift
distortions, and the dominant peculiar velocities in PT are co-linear
with density gradients.  The top right panel shows the bispectrum
quadrupole-to-monopole ratio, $R_{B}\equiv B_{s}^{(2)}/B_{s}^{(0)}$
for $r=k_{1}/k_{2}=10,2,1$ (top to bottom), in $\Omega=1$ (top set)
and $\Omega=0.3$ (bottom set) models, for $n=-2$ (solid) and $n=0$
(dotted).  As in the power spectrum case, $R_{B}$ increases with
$\Omega$, and it shows the expected configuration dependence; it is
relatively insensitive to the shape of the power spectrum.

The bottom right plot in Figure~1 quantifies how redshift distortions
alter the relation between galaxy and matter correlations.  For the
local, non-linear bias model of Eq.~(\ref{biasing}), the 3-point
hierarchical amplitude for galaxies satisfies $Q_{g}=(Q+\gamma)/b$ in
the absence of redshift distortions (Fry \& Gazta\~naga 1993, Fry
1994). As can be seen from the different terms in Eq.~(\ref{Bst}),
this simple relation no longer holds in redshift space.  The solid
curves show the exact PT result in redshift space, while the dotted
curves correspond to (wrongly) assuming that $Q_{g,s}=(Q_s+\gamma)/b$
holds in redshift space, i.e., that the operations of bias and
redshift-space mapping commute. Here we have set the linear bias to
$b=2$, the non-linear bias $\gamma=1/2,0,-1/2$ (from top to bottom),
and show configurations with $k_1/k_2 =2$, for a model with linear
spectral index $n=-2$. We see that the assumption that bias and
redshift distortions commute is accurate at about the 20\% level.

%
\subsection{Equilateral Configurations}
%

\begin{figure}[t!]
\centering
\centerline{\epsfxsize=18truecm\epsfysize=18truecm\epsfbox{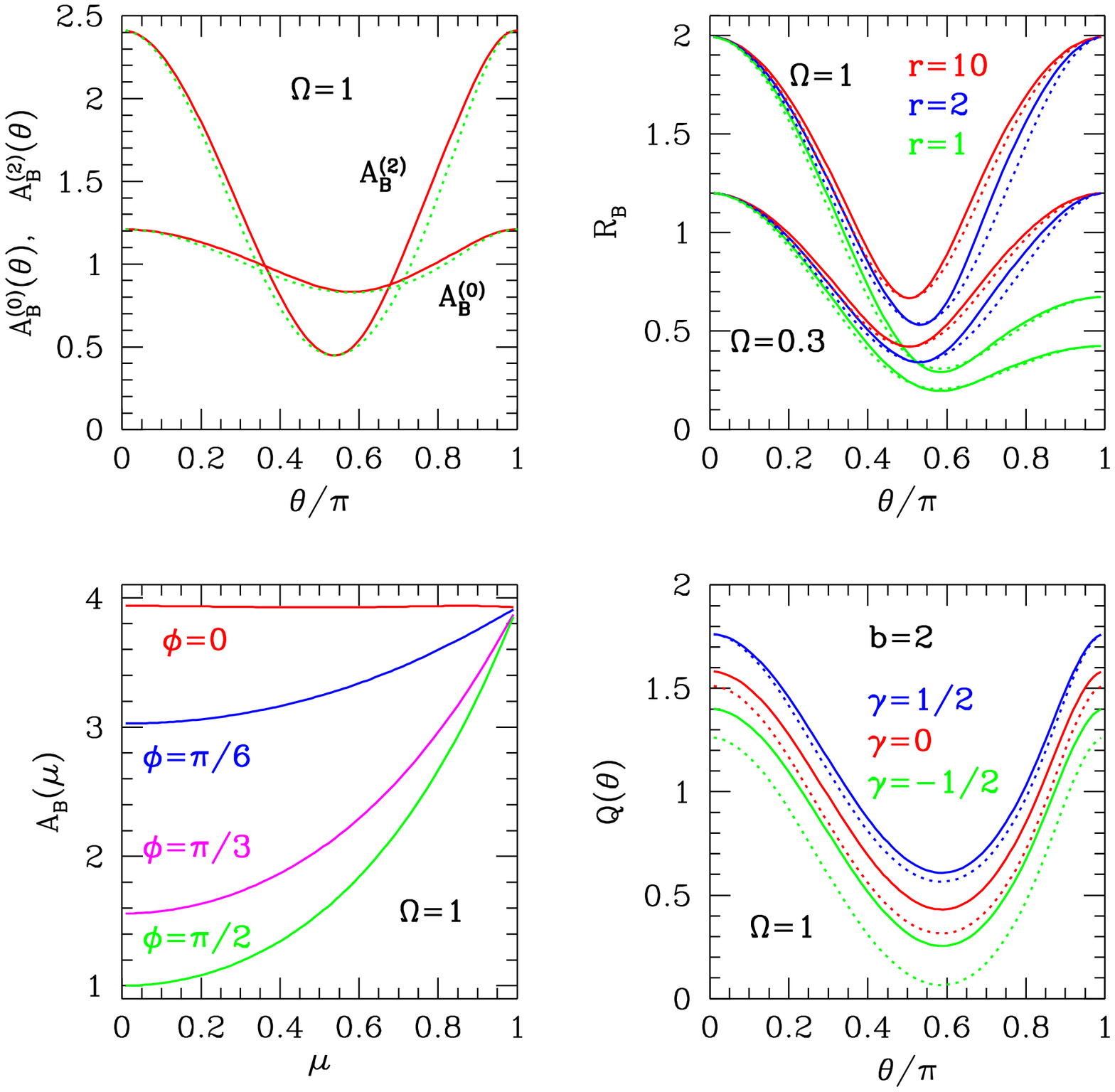}}
\caption{Top left panel shows the redshift-space tree-level bispectrum
multipoles, $A_B^{(\ell)} \equiv B_{s}^{(\ell)}/B$, as a function of
angle $\theta = \cos^{-1} \hat{\k}_1 \cdot \hat{\k}_2$, for
configurations with $k_1/k_2=2$ and scale-free initial spectra, $P(k)
\propto k^n$ ($n=-2$ solid, $n=0$ dotted). Top right panel shows the
bispectrum quadrupole-to-monopole ratio $R_B =
B_{s}^{(2)}/B_{s}^{(0)}$ for $r=k_1/k_2=10,2,1$ configurations as a
function of $\theta$ ($n=-2$ solid, $n=0$ dotted) for $\Omega=1$ and
$\Omega=0.3$. The bottom left plot shows the ratio of the
redshift-space to real-space bispectrum for equilateral triangles, as
a function of $\mu = \hat{\k}_1 \cdot \hat{z}$ for different azimuthal
angles $\phi$.  The bottom right plot shows the hierarchical amplitude
$Q_s$ for $k_1/k_2=2$ configurations with linear bias $b=2$, and
quadratic bias $\gamma=b_2/b=1/2,0,-1/2$ (top to bottom), for $n=-2$.
Solid curves represent the PT result, and dotted curves assume that
bias and redshift-space mapping commute.  }
\label{fig1}
\end{figure}

For equilateral triangle configurations ($k_1 = k_2 = k_3$),
the leading-order PT expression for the redshift-space bispectrum, $B_{s\ \rm
eq}$, is

\beqa
\label{beqs_munu}
B_{s\ \rm eq}(\mu,\nu)&=& [P_g(k)]^2 \Big(
	256 + 448\,\gamma  + 288\,\beta \,{{\mu }^2} + 
       224\,b\,\beta \,{{\mu }^2} + 448\,\beta \,\gamma \,{{\mu }^2} + 
       72\,{{\beta }^2}\,{{\mu }^4} + 168\,b\,{{\beta }^2}\,{{\mu }^4} + \nonumber \\
  & &
       84\,{{\beta }^2}\,\gamma \,{{\mu }^4} + 4\,{{\beta }^3}\,{{\mu }^6} - 
       7\,b\,{{\beta }^4}\,{{\mu }^8} + 288\,\beta \,{{\nu }^2} + 
       224\,b\,\beta \,{{\nu }^2} + 448\,\beta \,\gamma \,{{\nu }^2} - 
       288\,\beta \,{{\mu }^2}\,{{\nu }^2} - 
       \nonumber \\
  & &
       224\,b\,\beta \,{{\mu }^2}\,{{\nu }^2} + 
	144\,{{\beta }^2}\,{{\mu }^2}\,{{\nu }^2} + 
       336\,b\,{{\beta }^2}\,{{\mu }^2}\,{{\nu }^2} - 
       448\,\beta \,\gamma \,{{\mu }^2}\,{{\nu }^2} + 
       168\,{{\beta }^2}\,\gamma \,{{\mu }^2}\,{{\nu }^2} - 
       144\,{{\beta }^2}\,{{\mu }^4}\,{{\nu }^2} - \nonumber \\
  & &
       336\,b\,{{\beta }^2}\,{{\mu }^4}\,{{\nu }^2} - 
       24\,{{\beta }^3}\,{{\mu }^4}\,{{\nu }^2} - 
       168\,{{\beta }^2}\,\gamma \,{{\mu }^4}\,{{\nu }^2} + 
       24\,{{\beta }^3}\,{{\mu }^6}\,{{\nu }^2} + 
       35\,b\,{{\beta }^4}\,{{\mu }^6}\,{{\nu }^2} - 
       35\,b\,{{\beta }^4}\,{{\mu }^8}\,{{\nu }^2} + \nonumber \\
  & &
       72\,{{\beta }^2}\,{{\nu }^4} + 168\,b\,{{\beta }^2}\,{{\nu }^4} + 
       84\,{{\beta }^2}\,\gamma \,{{\nu }^4} - 
       144\,{{\beta }^2}\,{{\mu }^2}\,{{\nu }^4} - 
       336\,b\,{{\beta }^2}\,{{\mu }^2}\,{{\nu }^4} + 
       36\,{{\beta }^3}\,{{\mu }^2}\,{{\nu }^4} - \nonumber \\
  & &        
       168\,{{\beta }^2}\,\gamma \,{{\mu }^2}\,{{\nu }^4} +
	72\,{{\beta }^2}\,{{\mu }^4}\,{{\nu }^4} + 
       168\,b\,{{\beta }^2}\,{{\mu }^4}\,{{\nu }^4} - 
       72\,{{\beta }^3}\,{{\mu }^4}\,{{\nu }^4} - 
       21\,b\,{{\beta }^4}\,{{\mu }^4}\,{{\nu }^4} + 
       84\,{{\beta }^2}\,\gamma \,{{\mu }^4}\,{{\nu }^4} + \nonumber \\
  & &
       36\,{{\beta }^3}\,{{\mu }^6}\,{{\nu }^4} + 
       42\,b\,{{\beta }^4}\,{{\mu }^6}\,{{\nu }^4} - 
       21\,b\,{{\beta }^4}\,{{\mu }^8}\,{{\nu }^4} - 
       63\,b\,{{\beta }^4}\,{{\mu }^2}\,{{\nu }^6} + 
       189\,b\,{{\beta }^4}\,{{\mu }^4}\,{{\nu }^6} - \nonumber \\
  & &
       189\,b\,{{\beta }^4}\,{{\mu }^6}\,{{\nu }^6} + 
       63\,b\,{{\beta }^4}\,{{\mu }^8}\,{{\nu }^6} \Big)\ \times 3/(448\ b)
\eeqa
where $\nu\equiv\cos\phi$. For comparison, in real space, the PT
bispectrum is simply $B_{\rm eq}= (12/7) [P_g(k)]^2$ (Fry 1984).  The
bottom left panel in Fig.~\ref{fig1} shows the ratio of the redshift-
and real-space equilateral bispectra in leading order PT, $A_B \equiv
B_{s\ \rm eq}(\mu,\nu)/B_{\rm eq}$.  For $\mu=0$ and $\phi=\pi/2$, the
triangle lies in the plane perpendicular to the line-of-sight, and
there are no redshift distortions ($A_B=1$). The maximum distortion
occurs for $\mu=0$, $\phi=\pi$, with an enhancement factor $A_B =
189/48 =3.94$. A similar enhancement ($A_B = 1005/256=3.93$) is
obtained for the configuration with $\mu=1$, independent of $\phi$.

Averaging over $\phi$, the equilateral redshift-space bispectrum becomes

\beqa
B_{s\ \rm eq}(\mu)&=& [P_g(k)]^2\ \Big(
	4096 + 2304\,\beta  + 1792\,b\,\beta  + 432\,{{\beta }^2} + 
       1008\,b\,{{\beta }^2} + 7168\,\gamma  + 3584\,\beta \,\gamma  + 
       504\,{{\beta }^2}\,\gamma  + \nonumber \\
  & &
       2304\,\beta \,{{\mu }^2} + 
	1792\,b\,\beta \,{{\mu }^2} + 288\,{{\beta }^2}\,{{\mu }^2} + 
       672\,b\,{{\beta }^2}\,{{\mu }^2} + 216\,{{\beta }^3}\,{{\mu }^2} - 
       315\,b\,{{\beta }^4}\,{{\mu }^2} + 3584\,\beta \,\gamma \,{{\mu }^2} + 
       \nonumber \\
  & &
       336\,{{\beta }^2}\,\gamma \,{{\mu }^2} + 
	432\,{{\beta }^2}\,{{\mu }^4} + 1008\,b\,{{\beta }^2}\,{{\mu }^4} - 
       624\,{{\beta }^3}\,{{\mu }^4} + 819\,b\,{{\beta }^4}\,{{\mu }^4} + 
       504\,{{\beta }^2}\,\gamma \,{{\mu }^4} + \nonumber \\
  & &
       472\,{{\beta }^3}\,{{\mu }^6} - 413\,b\,{{\beta }^4}\,{{\mu }^6} - 
       203\,b\,{{\beta }^4}\,{{\mu }^8} 
  \Big)\ \times 3/(7168\ b)
\eeqa
Decomposing into Legendre polynomials, $B_{s\ \rm eq}(\mu) =
\sum_{\ell=0}^\infty\ B_{s\ \rm eq}^{(\ell)}\ P_{\ell}(\mu)$, and
defining the equilateral hierarchical amplitude in redshift-space,
$Q_{s\ {\rm eq}}^{(\ell=0)} \equiv B_{s\ \rm eq}^{(\ell=0)}/
3[P_s(k)]^2$, we obtain

\beq
\label{Qs0}
Q_{s\ {\rm eq}}^{(\ell=0)}=
\frac{5\ (2520+ 4410\,\gamma + 1890\,\beta + 2940\,\gamma\,\beta +
378\,\beta^2 + 441\,\gamma\,\beta^2 + 9\,\beta^3 + 1470\,b\,\beta
+ 882\,b\,\beta^2- 14\,b\,\beta^4 ) }{98\,b\, ( 15 + 10\,\beta +
3\,\beta^2 )^2}.
\eeq
This result shows once again that in redshift space, $Q_{s,g} \neq
(Q_{s} + \gamma)/b$ (although, as shown in Fig.~1, it is not a bad
approximation).  In the absence of bias ($b=1$, $\gamma =0$),
Eq.~(\ref{Qs0}) yields
\beq
\label{Qs0b0}
Q_{s\ {\rm eq}}^{(\ell=0)}=
{{5\,\left( 2520 + 3360\,f + 1260\,{f^2} + 9\,{f^3} - 14\,{f^4}
        \right) }\over
   {98\,{{\left( 15 + 10\,f + 3\,{f^2} \right) }^2}}},
\eeq
which approaches the real-space result $Q_{\rm eq}=4/7 =0.57$ in the
limit $f \sim \Omega^{0.6} \rightarrow 0$. On the other hand, for
$f=\Omega=1$, we have $Q_{s\ {\rm eq}}^{(0)}=0.464$: for these
configurations, the hierarchical 3-point amplitude is suppressed by
redshift distortions.  For the quadrupole-to-monopole ratio of $B_{\rm
eq}$ we have

\beq
\label{RBeq}
R_{\rm B\ eq}=
\frac{5\ ( 4158\,\beta + 3234\,\gamma\,\beta +
1188\,\beta^2 + 693\,\gamma\,\beta^2 + 33\,\beta^3 +3234\,b\,\beta +
2772\,b\,\beta^2 - 56\,b\,\beta^4 ) }{22 (2520+ 4410\,\gamma +
1890\,\beta + 2940\,\gamma\,\beta +
378\,\beta^2 + 441\,\gamma\,\beta^2 + 9\,\beta^3 + 1470\,b\,\beta +
882\,b\,\beta^2
- 14\,b\,\beta^4 ) }.
\eeq
For no bias and $\Omega=1$, this gives $R_{\rm B\
eq}=11329/31394=0.36$.  Eqs.~(\ref{Qs0}-\ref{RBeq}) provide two
constraints upon $(b,\beta,\gamma)$ that are independent of the power
spectrum. Thus, these relations for $Q_{s\ {\rm eq}}^{(\ell=0)}$ and
$R_{\rm B\ eq}$, together with the power spectrum
quadrupole-to-monopole ratio $R_P$ from Eq.~(\ref{PS_l}), could in
principle be inverted to obtain $\Omega$, $b$, and $b_2$ at large
scales, independent of the initial conditions. In practice, $R_P$ will
determine $\beta$, while $Q_{s\ {\rm eq}}^{(\ell=0)}$ and $R_{\rm B\
eq}$ then constrain the $b,\gamma$ parameter space. Moreover, $R_B$ is
very insensitive to $b$ and only mildly sensitive to $\gamma$, so that
there will typically be some residual degeneracy between $b$ and
$\gamma$.  To break this degeneracy, one could use the
shape-dependence of $B$, i.e., non-equilateral configurations (Fry
1994).

%
%
\section{Results: The Redshift-Space Bispectrum in Numerical Simulations}
%

%
%
\subsection{Numerical Simulations}
%

The numerical simulations used in this work correspond to
cluster-normalized cold dark matter (CDM) models run by the Virgo
collaboration (see e.g., Jenkins et al. 1998 for more details). We
present results for standard CDM (hereafter SCDM, with $\Omega=1$,
$h=0.5$) and ``Lambda'' CDM (hereafter $\Lambda$CDM, with
$\Omega=0.3$, $\Omega_\Lambda=0.7$, $h=0.7$), although we have also
done measurements in $\tau$CDM ($\Omega=1$ with power spectrum shape
parameter $\Gamma=0.21$) and open CDM ($\Omega=0.3$, $h=0.7$)
models. We found the results for SCDM and $\Lambda$CDM to be
representative of the full set of simulations. These simulations each
contain $256^3$ particles in a box $240 h^{-1}$ Mpc on a side. They
were run with an adaptive P$^3$M code (Couchman, Thomas \& Pearce,
1995; Pearce \& Couchman, 1997), and the initial conditions were set
using the Zel'dovich approximation on a ``glass'' (see, e.g. White
1996). We have also made measurements in a set of smaller simulations
run by the Hydra Consortium (Couchman, Thomas \& Pearce, 1995), with
very similar results.

We use the plane-parallel approximation to construct the
redshift-space maps, so that the periodic boundary conditions in the
simulation box are preserved by the redshift-space mapping. We can
therefore continue to use the FFT to obtain the density field in
Fourier space. This also has the advantage that our N-body
measurements are done using the same approximation as our perturbative
calculations.  We consider four different observer's directions, in
all cases misaligned with respect to the simulation box sides to avoid
spurious effects. Error bars in the plots denote the dispersion in the
values obtained for these four observers. To measure the bispectrum in
the numerical simulations, we use the method described in Scoccimarro
et al. (1998), developed by S. Colombi, with minor improvements such
as using TSC instead CIC to do the interpolations.

%
\subsection{Phenomenological Description of Non-Linear Redshift Distortions}
%

In order to describe the non-linear behavior of the redshift-space
bispectrum, we introduce a phenomenological model to take into account
the velocity dispersion effects and to quantify their importance. We
follow the standard approach in the literature (Peacock \& Dodds
1994), in which the non-linear distortions of the power spectrum in
redshift-space are written in terms of the linear squashing factor and
a suitable damping factor due to the pairwise-velocity distribution
function 

\beq
\label{Ppheno}
P_s(\k)= P_g(k)\ \frac{(1+\beta
\mu^2)^2}{[1+(k\mu\sigma_v)^2/2]^2} ~.
\eeq
Here $\sigma_v$ is a free parameter that characterizes the velocity
dispersion along the line-of-sight. This Lorentzian form of the
damping factor is motivated by empirical results showing an
exponential one-particle velocity distribution function (Park et
al. 1994); comparison with N-body simulations have shown it to be the
best phenomenological model so far (Cole, Fisher \& Weinberg 1995;
Hatton \& Cole 1998).  It is straightforward to obtain the multipole
moments of $P_s(\k)$ in this simple model (Cole, Fisher \& Weinberg
1995). In comparing with numerical simulations, we use the
quadrupole-to-monopole ratio statistic $R_P=a_2/a_0$ to fit
$\sigma_v$.

We propose a similar ansatz to model non-linear distortions of the
bispectrum:

\beq
\label{Bpheno}
B_s(\k_1,\k_2,\k_3)= \frac{B_s^{\rm PT}(\k_1,\k_2,\k_3)}{
\big[1+\alpha^2\ [(k_1\mu_1)^2 + (k_2\mu_2)^2 + (k_3\mu_3)^2]^2
\sigma_v^2/2\big]^2}, \eeq where $B_s^{\rm PT}(\k_1,\k_2,\k_3)$ is the
tree-level redshift-space bispectrum derived in the last section. Note
that we have written the triplet velocity dispersion along the
line-of-sight in terms of the pairwise velocity dispersion parameter
$\sigma_v$ and a constant $\alpha$ which reflects the configuration
dependence of the triplet velocity dispersion. As noted above,
$\sigma_v$ is determined from simulations solely using the power
spectrum ratio $R_P$; the parameter $\alpha$ is then fitted by
comparison with the monopole-to-quadrupole ratio of the equilateral
bispectrum, $R_B$, measured in the simulations.  The simpler
phenomenological model in which the supression factor in the
denominator in Eq.~(\ref{Bpheno}) is replaced by $1+ [(k_1\mu_1)^2 +
(k_2\mu_2)^2 + (k_3\mu_3)^2] \sigma_v^2/2$ does well in comparison
with the N-body results for configurations where $k_2/k_1=2$, but does
not reproduce the results of the equilateral quadrupole-to-monopole
ratio $R_B$. In what follows we find good fits to the measured
bispectrum using $\alpha \simeq 2$ for equilateral configurations and
$\alpha \simeq 3$ for $k_1/k_2=2$ configurations, independent of
cosmology (at redshift $z=0$). For $z=1$, we find that $\alpha=1$
(SCDM) and $\alpha=1.75$ ($\Lambda$CDM) reproduce the shape of $R_B$.
This is similar to the modelling of the equilateral three-point
function by Matsubara (1994), who found it necessary to have a triplet
velocity dispersion parameter $\sigma_{3} \approx 1.7 \sigma_{v}$ to
fit the N-body simulations.  We note here that we have not attempted
to do any best-fit procedure to obtain $\sigma_{v}$ and $\alpha$:
these values were simply obtained by inspecting by eye the results for
$R_P$ and the bispectrum.  We also note that eqn.~(\ref{Bpheno})
reduces to the perturbative expression $B_s^{\rm PT}$ at large scales
($k \rightarrow 0$), so it has the correct limiting behavior
[e.g. Eqs.~(\ref{Qs0}-\ref{RBeq})].

%
\subsection{Comparison with Numerical Simulations}
%

\begin{figure}[t!]
\centering
\centerline{\epsfxsize=18truecm\epsfysize=18truecm\epsfbox{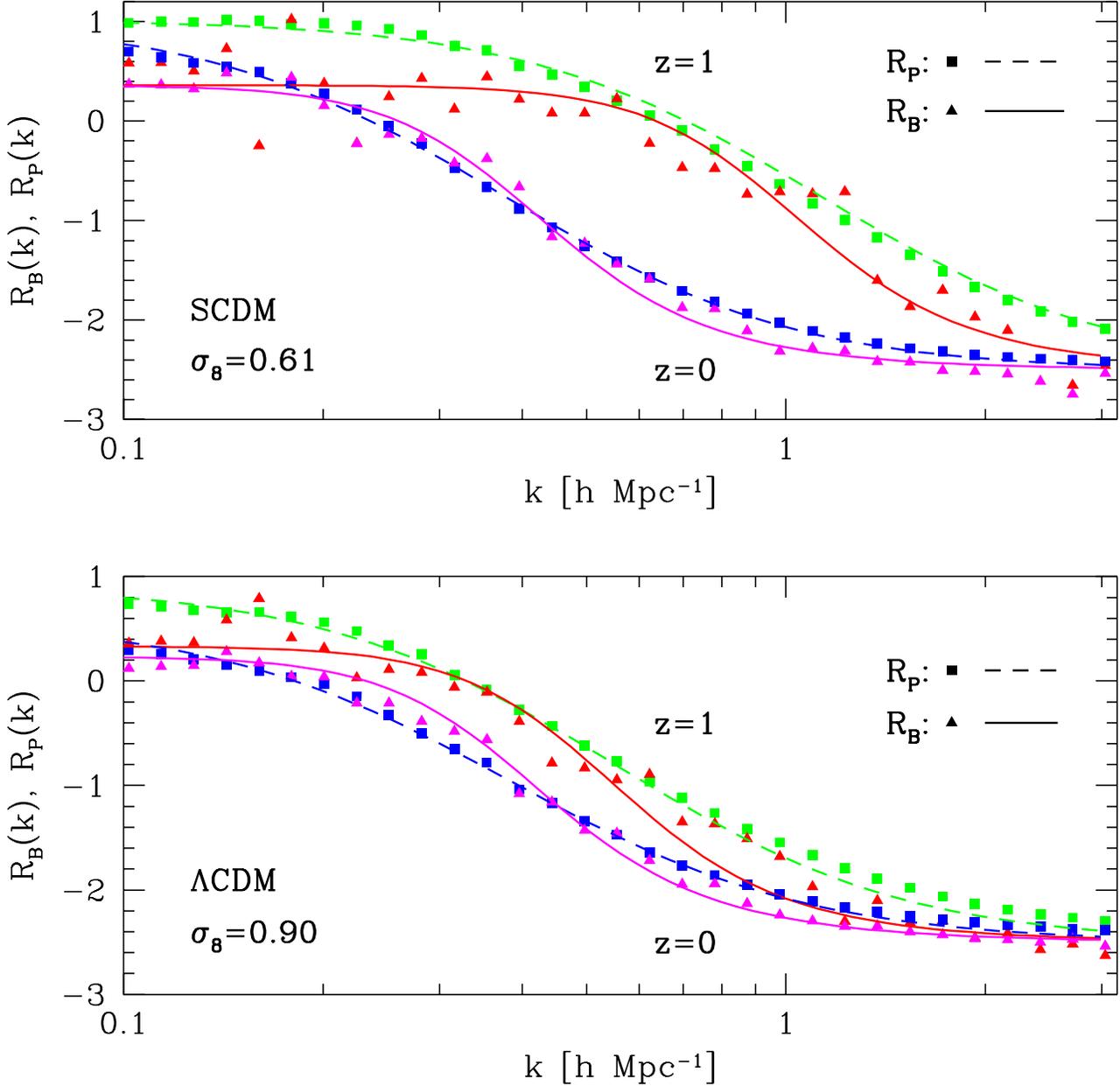}}
\caption{The quadrupole-to-monopole ratio for the redshift-space power
spectrum, $R_{\rm P}$, and equilateral bispectrum, $R_{\rm B}$. The
symbols correspond to AP$^3$M $256^3$ particle N-body simulations (see
text) at redshift $z=0,1$ averaged over four different observers, for
SCDM (top) and $\Lambda$CDM (bottom). Squares denote measured values
of $R_{\rm P}$, and triangles correspond to values of the ratio
$R_{\rm B}$. The dashed and solid curves show the predictions of PT
convolved with an exponential velocity dispersion model,
Eqs.~(\protect\ref{Ppheno}-\protect\ref{Bpheno}).  }
\label{fig2}
\end{figure}

Figure~\ref{fig2} shows the results from the phenomenological
redshift-distortion model and how it compares with N-body
simulations. Error bars have been suppressed for clarity but can be
guessed from the dispersion of the data points. For the SCDM model,
from $R_{\rm P}$ we obtain $\sigma_v=6$ (in units of H$_0$=100 $h$
km/s/Mpc) for $z=0$ and $\sigma_v=2$ for $z=1$. Using this result, we
can predict the equilateral bispectrum ratio $R_{\rm B}$ in the
phenomenological model by taking multipoles of Eq.~(\ref{Bpheno}). The
results are shown as the solid curves in Fig.~2, showing excellent
agreement with the N-body simulations. Similarly, for the $\Lambda$CDM
model, from fitting $R_{\rm P}$ we obtain $\sigma_v=5.5 $ for $z=0$
and $\sigma_v=4$ for $z=1$. The results for $R_{\rm B}$ are again in
very good agreement with the numerical simulations. Note the similar
behavior of $R_{\rm P}$ and $R_{\rm B}$ as a function of scale: the
FOG effect, which creates a deficit of pairs along the line-of-sight,
also causes a decrease of triplets compared to those in the
perpendicular direction. In this sense, it is a reassuring check to
note that in every case the zero-crossings of $R_{\rm P}$ and $R_{\rm
B}$ are very close to each other in scale.

The top left panel of Fig.~3 shows the power spectrum amplitude
$\Delta(k) = 4\pi k^3 P(k)$ in real (squares) and redshift (triangles)
space for SCDM, with the corresponding linear theory predictions. In
the figures, perturbative predictions are denoted by ``PT'' (real
space) and ``PTs'' (redshift space). At large scales, the squashing
effect boosts the power (Kaiser 1987), whereas at small scales the FOG
effect erases power in the monopole of the power spectrum. The top
right panel shows the equilateral hierarchical amplitude $Q$ in real
(squares) and redshift (triangles) space, with the corresponding
predictions of tree-level PT given by Eq.~(\ref{Qs0b0}) in redshift
space and $Q_{\rm eq}=4/7$ in real space. Note the similarity to the
power spectrum results: at small scales the FOG effect maintains the
redshift-space equilateral amplitude close to its perturbative value,
while in real space the amplitude grows by a large factor.  This
result, or the analog of this result for the skewness $S_3(R)$ (i.e.,
the fact that $S_3$ in redshift space is less scale-dependent than in
real space), has led previous authors to conclude that higher-order
correlations are `more hierarchical' in redshift-space than in real
space (Lahav et al. 1993; Matsubara 1994, Matsubara and Suto 1994;
Suto and Matsubara 1994; Bonometto et al. 1995; Ghigna et
al. 1996). However, this interpretation is in fact not correct: the
FOG effect at small scales decreases the equilateral amplitude with
respect to the real space value and shifts power to the co-linear
configurations. Therefore a strong configuration dependence of the
bispectrum is expected in redshift space, even when the real-space
correlations are well described by the hierarchical model (Scoccimarro
et al. 1998). This is exactly what we find in numerical simulations
(bottom right panels in Figs.~3 and~4), and what the model in
Eq.~(\ref{Bpheno}) predicts.

The bottom panels in Fig.~3 show a comparison of simulations to the
predictions of PT and the phenomenological model of Eq.~(\ref{Bpheno})
for configurations with $k_2/k_1=2$, for two different scales:
$k_1=10\ k_f$ (left) and $k_1=20\ k_f$ (right), where the fundamental
wavenumber of the simulation box is $k_f=2\pi/240\ h$Mpc$^{-1}$. In
the bottom left panel, where the amplitude $\lexp \Delta \rexp \approx
\Delta(k_2) = 2.5$ (thus somewhat beyond the weakly non-linear
regime), the tree-level PT calculation in real space (dashed curve)
predicts the form of the N-body results for $Q$ reasonably well,
although not to high accuracy.  However, the redshift-space PT
counterpart (dotted curve) does not provide a good description of the
simulation results for $Q_s$. This illustrates the point that PT tends
to break down on larger scales in redshift space than in real space.
On the other hand, the phenomenological model of Eq.~(\ref{Bpheno})
(solid curve) describes the N-body results for $Q_s$ very well. If one
were instead to use the redshift-space PT prediction [Eq.~(\ref{Bst})]
at these scales to probe the bias, one would mistakenly conclude that
the bias is $b=1.7$ instead of 1.

A similar set of plots is shown in Figure~4 for the $\Lambda$CDM
model, corresponding to $k_1/k_f=5,10,20$. Even at the largest scale
probed, corresponding to a wavelength $\lambda \simeq 50$ Mpc/$h$, the
tree-level PT prediction in redshift space is a poor fit to the
simulation results; using it, one would erroneously predict an
effective bias $b=1.4$, a quite significant discrepancy. The bottom
right panel nicely illustrates our point about correlations in the
strongly non-linear regime: whereas $Q$ is very close to hierarchical
(nearly configuration- and scale-independent), the redshift-space
amplitude $Q_s$ displays a strong configuration dependence. The
phenomenological model (solid curve) does an excellent job in
predicting $Q_s$, even at this stage of considerable non-linearity.

In a recent paper, Verde et al. (1998) independently proposed a
somewhat different phenomenological model to take into account
redshift-space distortions of the bispectrum, $\d_s(\k)=\d_s^{\rm
PT}(\k) / \sqrt{1+(k\mu\sigma_v)^2/2}$, which makes their damping
factor in Eq.~(\ref{Ppheno}) the square root of ours. We find that the
resulting power spectrum quadrupole-to-monopole ratio $R_{\rm P}$ does
not have the correct shape when compared to numerical simulations,
although it can be made to fit the low-$k$ behavior. If we fit
$\sigma_v$ to the simulation $R_P$ using their model, we find that the
predicted $Q_s^{(l=0)}$ does not provide a very accurate fit to the
simulation results. In particular, comparing their model to the
simulation results in Figs.~3 (bottom left) and~4 (top right) would
lead one to erroneously conclude that there is an effective bias of
$b=1.5$ and $b=1.4$, respectively.

The Verde et al. phenomenological model can however be improved as
follows (Heavens 1998, private communication).  The power spectrum
quadrupole-to-monopole ratio can be fixed by introducing a scale
dependence of the velocity dispersion parameter $\sigma_v \rightarrow
\sigma_v(k)$. Furthermore, using the non-linear power spectrum in the
perturbative formula for the bispectrum, Eq.~(\ref{Bs_def}), and
removing the wavevectors close to the line of sight leads to $Q_s$
values that agree with numerical simulations. However, this statistic
is not exactly equal to the monopole average $Q_s^{(l=0)}$, since some
configurations are not included in the averaging procedure.  We
conclude, therefore, that the phenomenological model of
Eq.~(\ref{Bpheno}) is the simplest way to accurately extract bias
parameters from redshift-space data.

%
%
\section{Conclusions and Discussion}
%
%

We have derived the fully non-linear expression for the density field
in redshift space in the plane-parallel approximation and applied this
result to obtain the redshift-space bispectrum to leading order in
PT. This derivation in Eulerian space is both simple and transparent
and can in fact be extended beyond leading order to understand the
large-scale FOG effect from a dynamical point of view (SCF). We
computed the multipole expansion of the bispectrum in redshift space,
in particular the monopole (Hivon et al. 1995) and quadrupole,
including the effects of non-linear biasing. These perturbative
results provide in principle a means of breaking the degeneracy
between bias and $\Omega$ present in measurements of the
redshift-space power spectrum on large scales.

We found that, even at relatively large scales (e.g., 50 $h^{-1}$Mpc)
the leading-order PT predictions in redshift space are not accurate
enough for detailed modelling and can lead to systematic errors in the
determination of the bias. This is similar to what happens with the
power spectrum: redshift-space statistics are more strongly affected
by non-linearities than their real-space counterparts.  In this
respect, it is interesting to note that even in the linear dynamics,
the exponential factor in Eq.~(\ref{d_s}) can lead to a FOG effect at
large scales (SCF). Therefore, the long range of the FOG effect seen
in numerical simulations, should not be attributed exclusively to
virialized clusters. This behavior agrees with the results of Taylor
\& Hamilton (1996) and Fisher \& Nusser (1996), who found that they
could qualitatively reproduce the shape of $R_{\rm P}$ seen in
numerical simulations by using the Zel'dovich approximation, although
quantitatively its use can lead to systematic errors (Hatton \& Cole
1998). The large-scale FOG effect has been recently connected to
redshift-space shell-crossing due to coherent infall by Hui, Kofman \&
Shandarin (1998).  All of these results strongly suggest the
possibility of extending the leading-order PT results for the power
spectrum and bispectrum to smaller scales by treating the
redshift-space mapping in Eq.~(\ref{d_s}) exactly and approximating
the dynamics using PT (SCF).

To correct for these non-linear effects, we have developed a
phenomenological model that convolves the PT results with a damping
factor due to velocity dispersion. Comparison with high-resolution
numerical simulations shows that this model, calibrated using the
power spectrum distortions, works very well and provides a
quantitative tool that can be applied to galaxy redshift surveys.

In the strongly non-linear regime, the reduced bispectrum is very
close to hierarchical (configuration-independent) in real space.
However, in redshift space we found that there is a strong
configuration dependence that reflects the velocity dispersion of
clusters. This helps to understand the deviations from the
hierarchical ansatz reported recently in the LCRS redshift-survey
three-point function (Jing \& B\"orner 1998).  For equilateral
configurations, on the other hand, the FOG effect suppresses the
scale-dependence of $Q_{s\ \rm eq}$ compared to that of $Q_{\rm eq}$,
and this can be misinterpreted as a sign of hierarchical correlations
in redshift space. A similar effect occurs with the skewness factor
$S_3$, which in redshift space shows a remarkable scale-independence
from weakly non-linear scales into the non-linear regime (Hivon et
al. 1995). Although $Q_{\rm eq}$ and $S_3$ in redshift space remain
very close to their PT values well into the non-linear regime, one
must be careful using the standard non-linear biasing prescriptions
beyond the regime of validity of tree-level PT, since
non-commutativity of redshift distortions and biasing may be more
complex than that exhibited at large scales.

\clearpage

\acknowledgments

We thank E.~Bertschinger, S.~Colombi, J.~Fry, A.~Hamilton, A.~Heavens,
L.~Hui, R.~Juszkiewicz, D.~Pogosyan, L.~Verde, and D.~Weinberg for
useful discussions. HMPC thanks CITA for hospitality during 1996--7
and frequent subsequent visits.  The numerical simulations analyzed in
this paper were carried out by the Virgo Supercomputing Consortium
(http://star-www.dur.ac.uk/~frazerp/virgo/virgo.html) using computers
based at the Max Plank Institut fur Astrophysik, Garching and the
Edinburgh Parallel Computing Centre.

%
%

\clearpage

\begin{figure}[t!]
\centering
\centerline{\epsfxsize=18truecm\epsfysize=18truecm\epsfbox{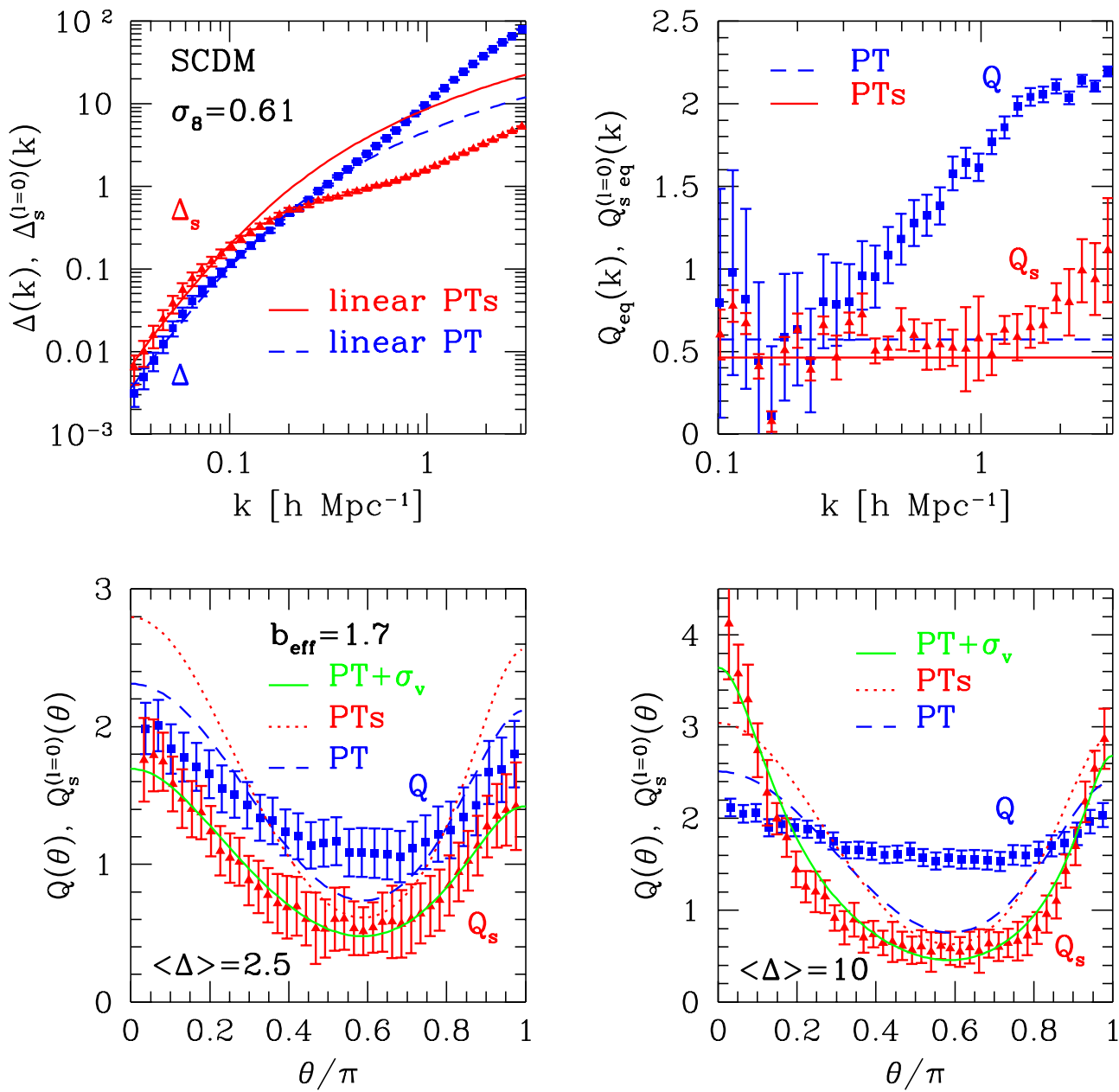}}
\caption{The top left panel shows the power spectrum amplitude,
$\Delta(k)\equiv 4 \pi k^3 P(k)$ in real (squares) and redshift
(triangles) space. The top right panel shows the equilateral
hierarchical amplitude $Q_{\rm eq}$ in real-space (squares) and the
monopole of the redshift-space amplitude, $Q_{s\ \rm eq}$ (triangles).
The bottom panels show the hierarchical amplitude $Q$ for $k_2/k_1= 2$
configurations, as a function of the angle $\theta$ between $\k_1$ and
$\k_2$, in real (squares) and redshift space (triangles), for two
different scales. In the bottom left plot, weakly non-linear
distortions decrease the configuration dependence of $Q_s$, similar to
the effects of bias in leading order PT.  At smaller scales (lower
right panel), the configuration dependence of $Q_s$ is greatly
enhanced, reflecting the velocity dispersion of virialized clusters.
}
\label{fig3}
\end{figure}

\begin{figure}[t!]
\centering
\centerline{\epsfxsize=18truecm\epsfysize=18truecm\epsfbox{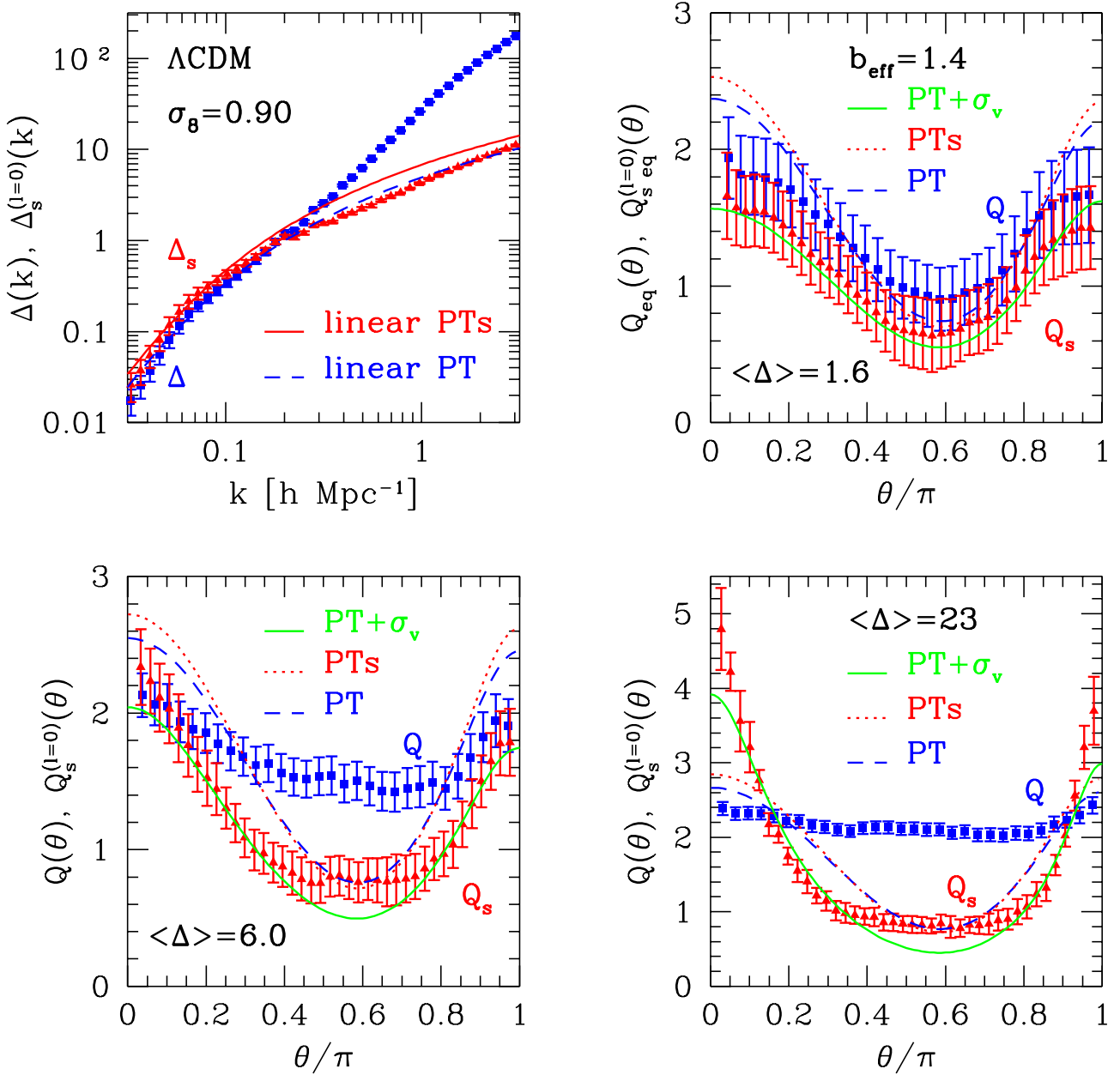 }}
\caption{Same as Fig~\ref{fig3}, for the $\Lambda$CDM model. In the
bottom right panel note the difference between the real- and
redshift-space hierarchical amplitudes. Whereas $Q$ in real-space is
approximately constant, its redshift-space counterpart depends
strongly on configuration. Therefore, the hierarchical ansatz does not
hold in redshift space.  }
\label{fig4}
\end{figure}

\end{document}